\documentclass[aps,prb,twocolumn,showpacs,tightenlines]{revtex4}

\begin{document}
\def \cT {{\cal T}}
\def \cI {{\cal I}}
\def \cP {{\cal P}}
\def \cG {{\cal G}}
\def\cB{{\cal B}}
\def \cU {{\cal U}}
\def \cV {{\cal V}}
\def \cF {{\cal F}}
\def \cT {{\cal T}}
\def \cH {{\cal H}}
\def \cA {{\cal A}}

\title{Landauer-B\"uttiker-type current formula for hybrid
mesoscopic systems}

\author{Z. Y. Zeng$^{1,2}$, F. Claro$^{2}$ and Baowen Li$^{1}$}
\affiliation{$^1$ Department of Physics, National University of
Singapore, 10 Kent Ridge Crescent, 119260 Singapore \\
$^2$Facultad de F\'isica, Pontificia Universidad Cat\'olica de
Chile, Casilla 306, Santiago 22, Chile}

\date{\today}
\draft

\begin{abstract}
A general Landauer-B\"uttiker-type current formula is derived,
which can be applied to the ferromagnet ({\bf F})/Normal metal
({\bf N})/superconductor ({\bf S}), {\bf F}/{\bf N}/{\bf N}, {\bf
N}/{\bf N}/{\bf S} and {\bf N}/{\bf N}/{\bf N} systems, even in
the presence of interactions in the central region.

\end{abstract}

\pacs{ 72.10.Bg,72.25.-b,74.25.Fy,73.40.-c}

\maketitle

\section{Introduction}

Electronic transport in mesoscopic systems or nanoscopic
structures has received extensive and intensive theoretical and
experimental attention \cite{Kouwenhoven}. In mesoscopic systems
the sample size is smaller than the phase coherent length,
electrons retain their phase when travelling through the sample.
In the ballistic limit, i.e., when the dimensions of the sample is
smaller than the mean free path, electrons can traverse the system
without any scattering. In contrast to macroscopic systems, the
conductance of mesoscopic systems is sample-specific, since
electron wavefunctions are strongly dependent on the form of the
boundary of the sample and the configuration of scatterers located
within the sample. To calculate the conductance of mesoscopic
samples, one should at first consider the wave nature of
electrons.
 The Boltzmann transport
equation\cite{Kittel} is obviously inappropriate, since the
preassumption that electrons can be viewed as classical particles
does not hold at a mesoscopic scale due to the Heinsenberg
uncertainty limitation. So one should resort to other theoretical
approaches such as  linear response theory. \cite{Kubo} However,
in fact, electronic transport in solids is equivalent to the wave
transmission of electrons through a generalized potential barrier,
which can be associated with a scattering matrix. In measuring the
conductance of a sample, one always connects the sample to some
contacts through perfect leads
 \cite{Datta}. In a two-terminal
setup (L=left, R=right), the Landauer-B\"uttiker
formula\cite{Buttiker}
 states that the current $\cI$   can be expressed as a convolution of the transmission
probability $\cT$ and the Fermi distribution function $f_{\alpha}$
($ \alpha=L,R$), i.e.,  $\cI= \frac{2e}{h}\int
\cT(\epsilon)[f_L(\epsilon)-f_R(\epsilon)]d\epsilon$. The
conductance $\cG$ in the linear response regime is
$\cG=\frac{2e^2}{h}\int \cT(\epsilon)(-\frac{\partial f}{\partial
\epsilon})d\epsilon$. Such a formulation seems more appealing
since the transport properties are encoded in the corresponding
transmission probability, which can be calculated by various
methods.

Due to the recent development in nanofabrication and material
growth technologies, several kinds of mesoscopic hybrid structures
have been realized experimentally. These nanoscale structures
include normal-metal/superconductor nanostructures ({\bf N}/{\bf
S})\cite{Poirier}, superconductor-insulator/superconductor
junctions ({\bf S}/{\bf I}/{\bf S})\cite{Morpurgo},
superconductor/quantum-dot/superconductor transistors ({\bf
S}/{\bf QD}/{\bf S})\cite{Tuominen}, normal metal/superconducting
quantum-dot/normal metal transistors ({\bf N}/{\bf SQD}/{\bf
N})\cite{Eiles}, ferromagnet/superconductor ({\bf F}/{\bf S})
contact \cite{Upadhyay}, superconductor/ferromagnet/superconductor
sandwich structure ({\bf S}/{\bf F}/{\bf S}) \cite{Lawrence} and
normal-metal/ferromagnetic-quantum-dot/normal metal ({\bf N}/{\bf
FQD}/{\bf N}) transistors \cite{Gueron}. In the presence of a
superconductor component, Andreev reflection\cite{Andreev}
dominates the transport process in the low bias case. The
imbalance between the spin-up and spin-down density of states at
the Fermi level for ferromagnetic materials introduces the
spin-dependent transport\cite {Prinz}. The conductance of a {\bf
F}/{\bf S} junction is predicted to be smaller or larger than the
{\bf N}/{\bf S} case, depending on the ratio between the spin-up
and spin-down density of states\cite {Jong}.

The nonequilibrium Green's function (NEGF) approach \cite {Haug}
has  proven to be a powerful technique to investigate the
transport problem in mesoscopic systems. Using the NEGF method,
Meir and Wingreen\cite{Meir} have derived a
Landauer-B\"uttiker-type formula for transport through an
interacting normal metal attached to two normal leads. Later, some
groups have applied the NEGF method to the {\bf N}/{\bf NQD}/{\bf
S}\cite{Sun} and {\bf S}/{\bf NQD}/{\bf S}\cite{Yeyati} cases. In
this paper, we extend the NEGF theory to a mesoscopic hybrid {\bf
F}/{\bf N}/{\bf S} structure, obtaining a Landauer-B\"uttiker-type
current formula, which allows us to investigate the spin-dependent
current and Andreev reflecting current in a unified way.

\section{Formulation of the Problem and Current Formulas}

We consider electron tunneling through a ferromagnet/normal
metal(semiconductor)/superconductor hybrid structure. Under the
mean-field approximation, the ferromagnet is characterized by a
molecular magnetic moment ${\bf M}$, making at an angle $\theta$
with the ${\bf F}/{\bf N}$ interface \cite{Slonczewski}, while the
BCS Hamiltonian is adopted for the superconductor, with an order
parameter $\Delta$ describing its energy gap \cite{Gennes}. In the
central region which contains a normal metal, we take into
consideration various kinds of couplings, such as the
electron-electron interaction, electron-phonon interaction, etc.
Then the hamiltonian can be written as
\begin{equation}
H=H_F+H_S+H_D+H_T,
\end{equation}
in which
\begin{eqnarray}
H_F &=&\sum_{k\sigma }(\epsilon _{k\sigma
}+\sigma M\cos \theta )f_{k\sigma }^{\dagger }f_{k\sigma
}+\sum_{k\sigma }M\sin \theta\nonumber \\ && f_{k\sigma }^{\dagger
}f_{k\stackrel{-}{\sigma }},   \\
H_S &=&\sum_{p\sigma }\epsilon _{p\sigma }s_{p\sigma }^{\dagger }s_{p\sigma
}+\sum_p\left[ \Delta ^{*}s_{p\uparrow }^{\dagger }s_{-p\downarrow
}^{\dagger }+\Delta s_{p\uparrow }s_{-p\downarrow }\right] , \\
H_C &=&\sum_{n\sigma }\epsilon _{n\sigma }c_{n\sigma }^{\dagger
}c_{n\sigma }+H_{int}(\{c_{n\sigma }^{\dagger}\},\{c_{n\sigma
}\})   \\
H_T &=&\sum_{kn;\sigma }\left[ T^L_{kn;\sigma }f_{k\sigma
}^{\dagger }c_{n\sigma }+T_{kn;\sigma }^{L*}c_{n\sigma }^{\dagger
}f_{k\sigma }\right] +\nonumber \\
& &\sum_{pn;\sigma }\left[ T^R_{pn;\sigma }s_{p\sigma }^{\dagger
}c_{n\sigma }+T_{pn;\sigma }^{R*}c_{n\sigma }^{\dagger }s_{p\sigma
}\right],
\end{eqnarray}
are the hamiltonian for ferromagnet, superconductor, central
normal metal and tunneling hamiltonian, respectively. In Eqs.
(2-5), $f(f^+)$, $s(s^+)$ and $c(c^+)$ represent the electron
annihilation (creation) operator of the ferromagnet,
superconductor and normal metal, respectively; $T^{L/R}$ denotes
the tunneling matrix between the normal metal and the left
ferromagnet (right superconductor), $H_{int}$ is the interaction
term in the central normal metal, which allows various kinds of
interaction to be included. Throughout this paper, we assume
$\Delta$ is real for convenience.

The current flowing into the central region from the left
ferromagnet can be evaluated from the time evolution of the total
electron number operator in the left lead \cite{Haug}
\begin{eqnarray}
J_L&=&-e<\frac{d N_L(t)}{dt}>= \frac{ie}{\hbar}<[N_L,H]>
      \nonumber \\
   &=&\frac{ie}{\hbar} \sum_{kn,\sigma}(T^L_{kn;\sigma}
      <f^\dagger_{k\sigma}(t)c_{n\sigma}(t)>-\nonumber \\ &&
      T^{L*}_{kn;\sigma}<c^\dagger_{n\sigma}(t)f_{k\sigma}(t)>)
      \nonumber\\
   &=&-\frac{e}{\hbar}Re{\sum\limits_{kn}^{i=1,3}\hat{\bf T}_{kn;ii}^{L\dagger}
\hat{\bf G}_{kn;ii}^<(t,t)}.
\end{eqnarray}
 Here we have expressed
 various kinds of  Green's functions in a
generalized Nambu representation, spanned in the spin-dependent
particle-hole space, in which the spin effect and Andreev
reflection
 are considered on the same footing and treated in a unified way. The
Green's functions in the generalized Nambu space are of the form
\begin{eqnarray}
\hat{\bf G}_{XY} ^{</>}(t,t')&=& \sum_{ij}\hat{\bf G}_{XiYj}
^{</>}(t,t') \nonumber \\
 &=&\pm i\sum_{ij}<{\bf
Y}_j^\dagger(t')/{\bf X}_i(t) \bigotimes \nonumber \\ &&
{\bf X}_i(t) /{\bf Y}_j^\dagger(t')>,\\
\hat{\bf G}_{XY} ^{r/a}(t,t')&=& \sum_{ij}\hat{\bf G}_{XiYj}
^{r/a}(t,t') \nonumber \\
&=&\mp i \sum_{ij}\theta(\pm t \mp
t')<[{\bf X}_i(t)\bigotimes{\bf Y}_j^\dagger(t')-\nonumber \\
& & {\bf Y}_j^\dagger(t')\bigotimes{\bf X}_i(t) >,
\end{eqnarray}
in which the four component of the spin-dependent particle-hole
vector reads
\begin{equation}
{\bf X}_i= \left(\matrix{ X_{i\uparrow } & X_{i \downarrow
}^{\dagger } & X_{i \downarrow } & X_{i\uparrow}^{\dagger } \cr }
 \right)^{\dagger}.
\end{equation}
The tunneling matrix takes the form in the generalized Nambu space
\begin{widetext}
\begin{equation}
\hat{\bf T}^L_{kn}/\hat{\bf T}^R_{pn}=\left( \matrix{
T^L_{kn;\uparrow}/T^R_{pn;\uparrow} & 0 & 0 & 0 \cr 0
&-T^{L*}_{kn;\downarrow}/T^{R*}_{pn;\downarrow} & 0 & 0 \cr 0 & 0
& T^L_{kn;\downarrow}/T^R_{pn;\downarrow} & 0 \cr 0 & 0 & 0
&-T^{L*}_{kn;\uparrow}/T^{R*}_{pn;\uparrow} \cr}\right).
\end{equation}
\end{widetext}

With the help of Dyson's equation, $\hat {\bf G}^<_{kn}$ can be
decoupled into the product of the unperturbed Green's function
$\hat{\bf g}_{kk}$ of the ferromagnet and the Green's function
$\hat{\bf G}_{mn}$ of the central normal metal regime in the
presence of elastic coupling to the outside world
\begin{eqnarray}
\hat{\bf G}_{kn}^<(t,t')&=&\sum_m\int dt_1 [\hat{\bf
g}_{kk}^<(t,t_1)\hat{\bf T}^L_{km}\hat{\bf G}_{mn}^r(t_1,t')+
\hat{\bf g}_{kk}^r(t,t_1) \nonumber \\ & &\hat{\bf
T}^L_{km}\hat{\bf G}_{mn}^<(t_1,t') ].
\end{eqnarray}
For convenience we introduce the linewidth matrices in the
generalized Nambu representation
\begin{eqnarray}
\hat{\bf \Gamma}^L(\omega)&=& 2\pi \hat{\bf T}_{kn}^{L\dagger}
\hat{\bf \rho}_{L}(\omega)\hat{\bf T}^L_{km} = i[
\hat{\bf \Sigma}^r_L(\omega)-\hat{\bf \Sigma}^a_L(\omega)],\\
\hat{\bf \Gamma}^R(\omega)&=&2\pi \hat{\bf T}_{pn}^{R\dagger}
\hat{\bf \rho}_{R}(\omega)\hat{\bf T}^R_{pm} = i[\hat{\bf
\Sigma}^r_R(\omega)-\hat{\bf \Sigma}^a_R(\omega)],
\end{eqnarray}
where (see Appendix)
\begin{eqnarray}
\hat{\bf \Sigma}^{r/a}_L(\omega)&=& \hat{\bf T}_{kn}^{L\dagger}
\hat{\bf g}_{ff}^{r/a}(\omega)\hat{\bf T}^L_{km} \\ \nonumber
 &=&\mp
\frac{i}{2} \left (\matrix{ \Gamma_{L+} & 0 & \Gamma_{L-} & 0 \cr
 0 &\Gamma'_{L+} & 0 & \Gamma_{L-}\cr
 \Gamma_{L-} & 0&\Gamma'_{L+} & 0 \cr
 0 & \Gamma_{L-} & 0 & \Gamma_{L+} \cr }\right), \\
\hat{\bf \Sigma}^{r/a}_R(\omega)&=& \hat{\bf T}_{pn}^{R\dagger}
\hat{\bf g}_{ss}^{r/a}(\omega)\hat{\bf T}^R_{pm}
\\ \nonumber
&=& \mp \frac{i}{2} \left(\matrix{
 \Gamma_{RD} & -\Gamma_{RND} & 0 & 0 \cr
 -\Gamma_{RND} & \Gamma_{RD} & 0 & 0 \cr
 0 & 0 & \Gamma_{RD} & \Gamma_{RND} \cr
  0 & 0 & \Gamma_{RND} & \Gamma_{RD} \cr}\right) ,
\end{eqnarray}
in which
\begin{eqnarray}
\Gamma_{L+}&=&\cos^2\frac{\theta}{2}\Gamma_{L\uparrow}+
\sin^2\frac{\theta}{2}\Gamma_{L\downarrow}, \\
\Gamma'_{L+}&=&\cos^2\frac{\theta}{2}\Gamma_{L\downarrow}+
\sin^2\frac{\theta}{2}\Gamma_{L\uparrow}, \\
\Gamma_{L-}&=& \frac{\sin \theta}{2}(\Gamma_{L\uparrow}-\Gamma_{L\downarrow}), \\
\Gamma_{RD}&=&\rho_S(\omega) \Gamma_R , \\
\Gamma_{RND}&=&\rho_S (\omega)\Gamma_R \frac{\Delta}{\omega} ,
\end{eqnarray}
with
\begin{eqnarray}
\Gamma_{L\uparrow}&=&2\pi \rho_{L\uparrow}T^{L*}_{kn} T^L_{km}, \\
\Gamma_{L\downarrow}&=&2\pi \rho_{L\downarrow}T^{L*}_{kn}
T^L_{km},\\
\Gamma_{R}&=&2\pi \rho_S^N T^{R*}_{pn}T^R_{Pm},\\
\rho_S(\omega)&=&\frac{|\omega|
\theta(|\omega|-\Delta)}{\sqrt{\omega^2-\Delta^2}}.
\end{eqnarray}
In the above equations, $\rho_{L\uparrow}$ and
$\rho_{L\downarrow}$ are the spin-up and spin-down density of
states in the ferromagnetic region, respectively, $\rho_S^N$ is
the normal density of states when the superconductor order
parameter $\Delta=0$, while $\rho_S$ is the dimensionless BCS
density of states of the superconductor.

 Substituting  Eq. (11) into (6),
we get the following compact form for the current
\begin{equation}
J_{L}= \frac{ie}{h}\sum\limits_{i=1,3}\int d\omega \{[\hat{\bf
G}_{cc}^{<}+ \hat {\bf f}_L(\hat{\bf
      G}_{cc}^r-\hat{\bf G}_{cc}^a)]\hat {\bf \Gamma}^{L} \}_{ii}. \\
\end{equation}
In deriving Eq. (25), we have used the fluctuation-dissipation
theorem $\hat{\bf g}^<=\hat{\bf f} ( \hat{\bf g}^a-\hat{\bf
g}^r)$, where $\hat{\bf f}$ is the Fermi distribution function
matrix in the generalized Nambu representation, given by
\begin{equation}
\hat{\bf f}_{L/R}=(\hat{f}_{L/R;ij}).
\end{equation}
Here $
\hat{f}_{L/R;ij}(x)=\delta_{ij}f_{L/R}(\omega+(-1)^i\mu_{L/R})$,
with  $\mu_{L/R}$ the chemical potential of the left
ferromagnet/right superconductor. Similarly, one can derive the
current flows into the right superconductor
\begin{equation}
J_{R}= \frac{ie}{h}\sum\limits_{i=1,3}\int d\omega \{[\hat{\bf
G}_{cc}^{<}+ \hat {\bf f}_R(\hat{\bf
      G}_{cc}^r-\hat{\bf G}_{cc}^a)]\hat {\bf \Gamma}^{R} \}_{ii}. \\
\end{equation}
Combining the Eqs. (26) and (27), we get the following current
formula( $\hat J=(\hat J_L-\hat J_R)/2$)
\begin{eqnarray}
J&=& \frac{ie}{2h}\sum\limits_{i=1,3}\int d\omega \{ (\hat {\bf
\Gamma}^{L}\hat {\bf f}_L- \hat {\bf \Gamma}^{R}\hat {\bf
f}_R)(\hat{\bf
      G}_{cc}^r-\hat{\bf G}_{cc}^a)+  \nonumber \\
      & &\hspace{3cm}
      (\hat {\bf
\Gamma}^{L}-\hat {\bf \Gamma}^{R})\hat{\bf G}_{cc}^{<} \}_{ii},
\end{eqnarray}
which is the central result of this work. It is the analogy of
Meir-wingreen\cite{Meir} formula in the presence of both
ferromagnetic and superconducting leads, providing a framework to
study transport in mesoscopic hybrid interacting structures.

In the absence of interactions within the central region, one has
the following Keldysh equation
\begin{eqnarray}
\hat {\bf G}_{cc}^<&=&\hat {\bf G}_{cc}^r\hat {\bf \Sigma}^<\hat
{\bf G}_{cc}^a,\\
\hat {\bf \Sigma}^<&=&\hat {\bf \Sigma}_L^<+\hat {\bf
\Sigma}_R^<=i(\hat {\bf f}_L \hat {\bf \Gamma}^L+\hat {\bf f}_R
\hat {\bf \Gamma}^R),
\end{eqnarray}
and the equality
\begin{eqnarray}
 \hat {\bf G}_{cc}^r-\hat {\bf
G}_{cc}^a&=&-i\hat {\bf G}_{cc}^r\hat {\bf \Gamma}\hat {\bf
G}_{cc}^a,\\
\hat {\bf \Gamma}&=&\hat {\bf \Gamma}^L+\hat {\bf \Gamma}^R.
\end{eqnarray}
With the help of Eqs. (29-32), Eq. (28) can be  simplified into
the following Landauer-B\"uttiker-type current formula for the
noninteracting {\bf F}/{\bf N}/{\bf S} mesoscopic hybrid structure
(setting $\mu_R=0$, $\mu_L=eV$ due to gauge invariance)
\begin{equation}
J=J^{NC}+J^A,
\end{equation}
in which
\begin{eqnarray}
J^{NC}&=& \frac{e}{h}\int d\omega [ f_L(\omega-e
V)-f_R(\omega)]\sum\limits_{i=1,3}\nonumber \\
 && [\hat{\bf
      G}_{cc}^r\hat{\bf \Gamma}^R\hat{\bf G}_{cc}^a
      \hat {\bf \Gamma}^L]_{ii}, \\
J^A&=& \frac{e}{h}\int d\omega [ f_L(\omega-e
V)-f_L(\omega+eV)]\sum\limits_{i=1,3}^{j=2,4}\hat{
      G}_{cc;ij}^r\nonumber \\ &&(\hat{\bf \Gamma}^L\hat{\bf G}_{cc}^a
      \hat {\bf \Gamma}^L)_{ji}.
\end{eqnarray}
From the expression (33) for the current, one sees that the
current comes from different physical process: (1) $J^{NC}$
includes (a) conventional electron tunneling current, (b)
tunneling current of an electron in the ferromagnet into the
superconductor either as a hole or as a cooper-pair by picking up
another electron; (2) $J^A$ is the Andreev reflection current,
representing an incident electron from the ferromagnet is
reflected as a hole backwards into the ferromagnet, with a
cooper-pair left in the superconductor. At zero temperature, when
$eV< \Delta$, $J^{NC}$ becomes zero ($\rho_S$ in $\hat{\bf
\Gamma}^R$ is zero when $eV< \Delta$) and the Andreev reflection
dominates the current. Obviously, the current has a strong
dependence on the polarization
$\cP=\frac{\rho_{L\uparrow}-\rho_{L\downarrow}}
{\rho_{L\uparrow}+\rho_{L\downarrow}}$ of the ferromagnet and the
direction $\theta$ of the magnetic moment ${\bf M}$, through the
coupling matrix $\hat {\bf \Gamma}^L$ (notice that the Green's
functions of the central region is also implicitly dependent on
$\hat {\bf \Gamma}^L$).  Equation (28) and Equations (33-35) can
be applied to the interacting and noninteracting mesoscopic  $\bf
F/ \bf N/\bf S$, $\bf N/ \bf N/\bf S$, $\bf F/ \bf N/\bf N$ and
$\bf N/ \bf N/\bf N$ structures, respectively.

\section {Examples}

In order to illustrate how  formulas (33-35) are applied to the
 $\bf F/ \bf N/\bf S$, $\bf N/ \bf N/\bf S$, $\bf F/
\bf N/\bf N$ and $\bf N/ \bf N/\bf N$ structures,  we discuss in
what follows a simple case, in which the interaction in the
central region is omitted, and only one level is relevant to the
transport,

   Consider the hamiltonian of the central region,
   $H_C=\epsilon_{c}
   c^\dagger c$. The unperturbed retarded/advanced Green's
   function $\hat {\bf g}^{r/a}_{cc}$
   in the $4\times
   4$ spin-dependent particle-hole space is then
   \begin{widetext}
   \begin{eqnarray}
   (\hat{\bf g}_{cc}^{r/a})^{-1}
 =\left (\matrix{
 \omega-\epsilon_{c} \pm i0^+ & 0 & 0
  & 0 \cr
 0 &
 \omega+\epsilon_{c} \pm i0^+& 0 & 0 \cr
 0 & 0&
 \omega-\epsilon_{c}\pm i0^+ & 0 \cr
 0 & 0 & 0 &
 \omega+\epsilon_{c} \pm i0^+  \cr }\right).
\end{eqnarray}
\end{widetext}
From Dyson's equation, $\hat {\bf G}^{r/a}=\hat{\bf
g}^{r/a}+\hat{\bf g}^{r/a}\hat{\bf \Sigma}^{r/a}\hat {\bf
G}^{r/a}$, one has for $\hat {\bf G}^{r/a}_{cc}$ the expression
\begin{eqnarray}
\hat {\bf G}^{r/a}_{cc}&=&(adj(\hat {\bf A})_{ij})det(\hat {\bf A})^{-1}, \\
\nonumber \hat {\bf A}&=&[(\hat{\bf g}^{r/a}_{cc})^{-1}-\hat{\bf
\Sigma}^{r/a}]^{-1},
\end{eqnarray}
where $adj(X)$ denotes the adjoint of $X$.  Note that $\hat {\bf
G}^{r/a}_{cc}$ is  symmetric , i.e., $\hat {\bf
G}^{r/a}_{cc;ij}=\hat {\bf G}^{r/a}_{cc;ji}$, as seen from its
definition or Eq. (36).

\subsection{current through a \bf F/\bf N/\bf S structure}

For simplicity, we just consider here $\theta=0$. Then the
self-energy matrix $\hat {\bf \Sigma}^{r/a}_L$ as well as the
linewidth matrix  $\hat {\bf \Gamma}^L$ is diagonal, and $\hat
{\bf G}^{r/a}_{cc}$ becomes
\begin{eqnarray}
  \hat {\bf G}^{r/a}_{cc}
 =\left (\matrix{ \hat {\bf G}_1^{r/a}
  &\hat{\bf 0} \cr
 \hat{\bf 0} & \hat {\bf G}_2^{r/a} \cr }\right),
\end{eqnarray}
where
\begin{widetext}
\begin{eqnarray}
  \hat {\bf G}_1^{r/a}&=&B_1^{-1}\left (\matrix{
 \omega+\epsilon_{c}\pm i(\Gamma_{L\downarrow}+ \Gamma_{RD}) &
 \mp i\Gamma_{RND}\cr
\mp i\Gamma_{RND} &
 \omega-\epsilon_{c}\pm i(\Gamma_{L\uparrow}+\Gamma_{RD})\cr }\right) \\
  \hat {\bf G}_2^{r/a}&=& B_2^{-1}\left (\matrix{
 \omega+\epsilon_{c}\pm i(\Gamma_{L\uparrow}+\Gamma_{RD}) & \pm i\Gamma_{RND}\cr
\pm i\Gamma_{RND} &
 \omega-\epsilon_{c} \pm i(\Gamma_{L\downarrow}+\Gamma_{RD})\cr }\right)
 \end{eqnarray}
 \end{widetext}
 with
 \begin{eqnarray*}
 B_1&=&det(\hat {\bf G}_1^{r/a}), \\
B_2&=&det(\hat {\bf G}_2^{r/a}).
\end{eqnarray*}
Then the current tunneling through the $\bf F/\bf N/{\bf S}$
structure is
\begin{widetext}
\begin{eqnarray}
J^{NC}&=& \frac{e\Gamma_{R}\rho_S}{h}\int d\omega [ f_L(\omega-e
V)-f_R(\omega)]\Big\{\Gamma_{L\uparrow}[\sum\limits_{i=1,2} |\hat
G^r_{cc; 1i}|^2-2\frac{\Delta}{\omega}Re(\hat G^r_{cc; 11}\hat
G^{r*}_{cc; 12})]+\cr & &\hspace{2cm}
\Gamma_{L\downarrow}[\sum\limits_{i=3,4} |\hat G^r_{cc;
3i}|^2+2\frac{\Delta}{\omega}Re(G^r_{cc;
33}G^{r*}_{cc; 34})]\Big \}, \\
J^{A}&=& \frac{e\Gamma_{L\uparrow}\Gamma_{L\downarrow}}{h}\int
d\omega [ f_L(\omega-e V)-f_L(\omega+e V)][ |\hat G^r_{cc;
12}|^2+|\hat G^r_{cc; 34}|^2].
\end{eqnarray}
\end{widetext}
The physical implications of Equations (41) and (42) are apparent.
$J^{NC}$ is directly proportional to the spin-dependent tunneling
matrix $\Gamma_{\sigma}$ due to the coupling between the central
region and the ferromagnet, and $\Gamma_{R}$ as well as $\rho_S$
for the superconductor. The Andreev current is proportional to the
spin-up and spin-down tunneling matrix $\Gamma_{L\uparrow}$ and
$\Gamma_{L\downarrow}$, which means that an up(down)-spin
electron(hole) incident from the ferromagnet will be reflected in
the interface of $\bf N/\bf S$ and re-enter the ferromagnet as a
down(up)-spin hole(electron). In contrast to the $\bf N/\bf N/\bf
S$ structure (below), the Andreev current is strongly dependent on
the polarization of the ferromagnet
$\cP=\frac{\rho_{L\uparrow}-\rho_{L\downarrow}}
{\rho_{L\uparrow}+\rho_{L\downarrow}}$ via the factor
$\Gamma_{L\uparrow} \Gamma_{L\downarrow}$. When the ferromagnet is
fully polarized, i.e., $\cP=1$ or $\cP=-1$, the Andreev current
will be zero since no state is available for the reflected
particle with reverse spin.

\subsection{current through a \bf N/\bf N/\bf S structure}

 When the molecule magnetic moment $\bf M$ is set to zero,
 the ferromagnet becomes a normal metal
structure. The spin-up and spin-down density of states
$\rho_{L\uparrow}$ and $\rho_{L\downarrow}$ will be the same. Then
$\hat \Gamma_{L\uparrow}=\hat \Gamma_{L\downarrow}=\hat
\Gamma_{L}$, and $\hat G^r_{cc; 11}=\hat G^r_{cc; 33}$, $\hat
G^r_{cc; 22}=\hat G^r_{cc; 44}$, and $\hat G^r_{cc; 12}=\hat
G^r_{cc; 21}=-\hat G^r_{cc; 34}=- \hat G^r_{cc; 43}$. Therefore
the current through the $\bf N/\bf N/\bf S$ system turns into
\begin{eqnarray}
J^{NC}&=& \frac{2e\Gamma_{R}\Gamma_{L}\rho_S}{h}\int d\omega [
f_L(\omega-e V)-f_R(\omega)]\nonumber \\
& &\Big\{[\sum\limits_{i=1,2} |\hat G^r_{cc;
1i}|^2-2\frac{\Delta}{\omega}Re(\hat G^r_{cc; 11}\hat G^{r*}_{cc;
12})], \\
J^{A}&=& \frac{2e\Gamma_{L}^2}{h}\int d\omega [ f_L(\omega-e
V)-f_L(\omega+e V)] \nonumber \\ & &
|\hat G^r_{cc; 12}|^2,
\end{eqnarray}
which is the same as the results for the $\bf N/\bf NQD/\bf S$
hybrid structure derived by Sun and co-workers \cite{Sun} and has
been  studied in detail.

\subsection{current through a \bf F/\bf N/\bf N structure}

When the order parameter $\Delta=0$, the superconductor can be
viewed as a normal metal. In this case, one can observe that
$\rho_S=1$ and all the non-diagonal elements of the full Green's
function $\hat {\bf G}^{r/a}_{cc}$ of the central region are zero.
Then the Andreev tunneling current $J^A=0$, and $J^{NC}$ is
simplified as
\begin{eqnarray}
J=J^{NC}&=& \frac{e\Gamma_{R}}{h}\int d\omega [ f_L(\omega-e
V)-f_R(\omega)][\Gamma_{L\uparrow} |\hat G^r_{cc; 11}|^2\nonumber
\\
& & +\Gamma_{L\downarrow} |\hat G^r_{cc; 33}|^2].
\end{eqnarray}
Since, for normal metal, one can write
$\Gamma_{R}=(\Gamma_{R\uparrow}+\Gamma_{R\downarrow})/2$, then one
gets the following expression for the current
\begin{eqnarray}
J&=&J_{\uparrow \uparrow}+J_{\downarrow \downarrow}+J_{\uparrow
\downarrow}+J_{\downarrow \uparrow}; \cr
J_{\uparrow \uparrow}&=&
\frac{e\Gamma_{L\uparrow}\Gamma_{R\uparrow}}{2h}\int d\omega [
f_L(\omega-e V)-f_R(\omega)] |\hat G^r_{cc; 11}|^2,\cr
J_{\downarrow \downarrow}&=&
\frac{e\Gamma_{L\downarrow}\Gamma_{R\downarrow}}{2h}\int d\omega [
f_L(\omega-e V)-f_R(\omega)]|\hat G^r_{cc; 33}|^2,\cr
 J_{\uparrow \downarrow}&=&
\frac{e\Gamma_{L\uparrow}\Gamma_{R\downarrow}}{2h}\int d\omega [
f_L(\omega-e V)-f_R(\omega)]|\hat G^r_{cc; 11}|^2,\cr
J_{\downarrow \uparrow}&=& \frac{e\Gamma_{L\downarrow}
\Gamma_{R\uparrow}}{2h}\int d\omega [ f_L(\omega-e
V)-f_R(\omega)]\nonumber \\
& &|\hat G^r_{cc; 33}|^2.
\end{eqnarray}
The physical meaning of Eq. (46) is very clear. The current into
the right normal region comes from two kinds of physical process:
one is that an electron exits from the ferromagnet and enters the
central normal region, and without spin-flip it tunnels into the
right normal metal; the other is the spin-flipped tunneling
process, the up(down)-spin electron leaving the ferromagnet into
the central normal metal, and after spin-flip, it gets into the
right normal region with down(up) spin.

\subsection{current through a \bf N/\bf N/\bf N structure}

Taking advantage of the results for an  $\bf F/\bf N/\bf N$
structure, if one further assumes
$\Gamma_{L\uparrow}=\Gamma_{L\downarrow}$, the $\bf F/\bf N/\bf N$
structure then turns into the $\bf N/\bf N/\bf N$ structure, and
one easily recovers from Eq. (46) the well-known expression
\cite{Jauho}
\begin{eqnarray}
J=J^{NC}&=& \frac{2e\Gamma_{L}\Gamma_{R}}{h}\int d\omega [
f_L(\omega-e V)-f_R(\omega)]|\hat G^r_{cc; 11}|^2 \nonumber \\
&=&
\frac{2e}{h}\int d\omega [ f_L(\omega-e
V)-f_R(\omega)]\nonumber \\
& &\hspace{1.5cm}
\frac{\Gamma_{L}\Gamma_{R}}{(\omega-\epsilon_c)^2+
(\Gamma_{L}+\Gamma_{R})^2/4}.
\end{eqnarray}
One sees from Eq. (47) that the transmission probability
$\cT\propto \frac{\Gamma_{L}\Gamma_{R}}{(\omega-\epsilon_c)^2+
(\Gamma_{L}+\Gamma_{R})^2/4}$ is of the Breit-Wigner form.

\section{Concluding Remarks}

We have given a general Landauer-B\"uttiker current formula, which
can be applied to the $\bf F/ \bf N/\bf S$, $\bf N/ \bf N/\bf S$,
$\bf F/ \bf N/\bf N$ and $\bf N/ \bf N/\bf N$ structures, even in
the presence of interactions within the central region. However,
one finds that one usually can not divide the current into the
spin-up and spin-down current parts. Then the polarization in the
presence of a ferromagnetic and a superconducting component can
not be defined by the current polarization. The reason is that the
spin-up/down current can not be conserved during the transport
process, due to the spin flip in the central region and Andreev
reflection at the $\bf N/\bf S$ interface. For example, a spin-up
electron from the ferromagnet tunnels through the barrier into the
central region, while another electron with spin-down will
probably enter the superconductor. It is obvious in such case that
the spin-up/down current is not conserved.

\begin{center}
{\bf ACKNOWLEDGMENT}
\end{center}
This work was supported by a \{Singapore\}, C\'atedra Presidencial
en Ciencias, FONDECYT 1990425, Chile and NSF grant No. 53112-0810
of Hunan Normal University,China (ZYZENG).

\begin{center}
{\bf Appendix}
\end{center}

In this Appendix we derive the retarded(advanced) self energy
matrices $\hat{\bf \Sigma}^{r/a}_L$ and $\hat{\bf
\Sigma}^{r/a}_L$, due to the coupling of electrons in the central
normal metal to the left ferromagnet and the right superconductor.
 The key step is to get first the retarded(advanced) Green's
function for the ferromagnet and the superconductor: $\hat{\bf
g}^{r/a}_{ff}$ and $\hat{\bf g}^{r/a}_{ss}$. The most convenient
way is to diagonalize the hamiltonian of the ferromagnet and
superconductor. We first calculate the Green's function for the
ferromagnet.  Applying the following Bogoliubov-Valatin
transformation \cite{Bogoliubov} to the ferromagnet hamiltonian
$H_F$
\begin{equation}
f_{k\sigma}=\cos (\theta/2) a_{k\sigma}-\sigma \sin(\theta/2) a_{k
\stackrel{-}{\sigma}},
\end{equation}
 one has
\begin{equation}
H'_{F} = \sum_{k\sigma}(\epsilon_{k\sigma}+\sigma M)
a_{k\sigma}^{\dagger }a_{k\sigma}.
\end{equation}
Defining the following retarded(advanced) Green's function
\begin{eqnarray}
g^{r/a}_{a\sigma;e}(t-t')&=&\mp i\theta(\pm t \mp t')\sum_{k}
 <{a_{k\sigma}(t),
a_{k\sigma}^{\dagger}(t')}> \nonumber\\
 &=& \mp i\theta(\pm t \mp
t')\sum_{k}e^{-i(\epsilon _{k\sigma }+\sigma M)(t-t')
/\hbar}, \\
g^{r/a}_{a\sigma;h}(t-t')&=&\mp i\theta(\pm t \mp t')\sum_{k}
<{a_{k\sigma} (t),a_{k\sigma}^{\dagger}(t')}> \nonumber\\
 &=& \mp
i\theta(\pm t \mp t')\sum_{k}e^{i(\epsilon _{k\sigma }+\sigma
M)(t-t') /\hbar},
\end{eqnarray}
one finally gets for the ferromagnetic lead  the retarded
(advanced) Green's functions
\begin{equation}
\hat {\bf g}_{ff}^{r/a}(t-t')=\sum_{k} \hat{\bf
g}^{r/a}_{fkfk}(t-t') =(\hat {g}_{ff;mn}^{r/a}(t-t')),
\end{equation}
where
\begin{eqnarray}
\hat {g}_{ff;11}^{r/a}(t-t')&=&\cos^2\frac{\theta}{2}g^{r/a}
_{a\uparrow;e}(t-t')+\nonumber \\ && \hspace{2cm}
\sin^2\frac{\theta}{2}g^{r/a}_{a\downarrow;e}(t-t'),\\
\hat {g}_{ff;22}^{r/a}(t-t')&=&\cos^2\frac{\theta}{2}g^{r/a}
_{a\downarrow;h}(t-t')+\nonumber \\ &&\hspace{2cm}
\sin^2\frac{\theta}{2}g^{r/a}_{a\uparrow;h}(t-t'),\\
\hat {g}_{ff;33}^{r/a}(t-t')&=&\cos^2\frac{\theta}{2}g^{r/a}
_{a\downarrow;e}(t-t')+\nonumber \\ &&\hspace{2cm}
\sin^2\frac{\theta}{2}g^{r/a}_{a\uparrow;e}(t-t'),\\
\hat {g}_{ff;44}^{r/a}(t-t')&=&\cos^2\frac{\theta}{2}g^{r/a}
_{a\uparrow;h}(t-t')+\nonumber \\ &&\hspace{2cm}
\sin^2\frac{\theta}{2}g^{r/a}_{a\downarrow;h}(t-t'),\\
\hat {g}_{ff;13}^{r/a}(t-t')&=&\hat
{g}_{ff;31}^{r/a}(t-t')\nonumber \\ &=&
\frac{\sin{\theta}}{2}(g^{r/a}_{a\uparrow;e}(t-t')-g^{r/a}
_{a\downarrow;e}(t-t')),\\
\hat {g}_{ff;24}^{r/a}(t-t')&=&\hat {g}_{ff;42}^{r/a}(t-t')
\nonumber \\ &=&\frac{\sin{\theta}}{2}
(g^{r/a}_{a\uparrow;h}(t-t')-g^{r/a}_{a\downarrow;h}(t-t')),\\
\hat {g}_{ff;12}^{r/a}(t-t')&=&\hat {g}_{ff;14}^{r/a}(t-t') =\hat
{g}_{ff;21}^{r/a}(t-t')\nonumber \\ &=&
\hat {g}_{ff;23}^{r/a}(t-t')=0,\\
\hat {g}_{ff;32}^{r/a}(t-t')&=&\hat {g}_{ff;34}^{r/a}(t-t')= \hat
{g}_{ff;41}^{r/a}(t-t')\nonumber \\ &=&\hat
{g}_{ff;43}^{r/a}(t-t')=0.
\end{eqnarray}
The sum over $k$ can be transformed into an integral, i.e.,
$\sum_{k}\rightarrow \int d\epsilon \rho_{L\sigma}$. Then one
arrives at
\begin{eqnarray}
\hat {g}_{ff;11}^{r/a}(t-t')&=&\mp i\delta(t-t')
(\cos^2\frac{\theta}{2}\rho_{L\uparrow}+\nonumber \\
&&\hspace{1cm}
\sin^2\frac{\theta}{2}\rho_{L\downarrow}),\\
\hat {g}_{ff;22}^{r/a}(t-t')&=&\mp i\delta(t-t')
(\cos^2\frac{\theta}{2}\rho_{L\downarrow}+\nonumber \\
&&\hspace{1cm}
\sin^2\frac{\theta}{2}\rho_{L\uparrow}),\\
\hat {g}_{ff;33}^{r/a}(t-t')&=&\mp i\delta(t-t')
(\cos^2\frac{\theta}{2}\rho_{L\downarrow}+\nonumber \\
&&\hspace{1cm}
\sin^2\frac{\theta}{2}\rho_{L\uparrow}),\\
\hat {g}_{ff;44}^{r/a}(t-t')&=&\mp i
\delta(t-t')(\cos^2\frac{\theta}{2}\rho_{L\uparrow}+\nonumber \\
&&\hspace{1cm}
\sin^2\frac{\theta}{2}\rho_{L\downarrow}),\\
\hat {g}_{ff;13}^{r/a}(t-t')&=&\hat {g}_{ff;31}^{r/a}(t-t')=
\hat {g}_{ff;24}^{r/a}(t-t')\nonumber \\
& =&\hat {g}_{ff;42}^{r/a}(t-t')
   =\mp i\delta(t-t')\nonumber \\&& \hspace{1cm}
   \frac{\sin{\theta}}{2}
(\rho_{L\uparrow}-\rho_{L\downarrow}),\\
\hat {g}_{ff;12}^{r/a}(t-t')&=&\hat {g}_{ff;14}^{r/a}(t-t') =\hat
{g}_{ff;21}^{r/a}(t-t')\nonumber \\ &=&
\hat {g}_{ff;23}^{r/a}(t-t')=0, \\
\hat {g}_{ff;32}^{r/a}(t-t')&=&\hat {g}_{ff;34}^{r/a}(t-t')= \hat
{g}_{ff;41}^{r/a}(t-t')\nonumber \\ &=&\hat
{g}_{ff;43}^{r/a}(t-t')=0.
\end{eqnarray}
The matrix product $\hat{\bf L}_{kn}^\dagger \hat{\bf
g}_{ff}^{r/a}(t-t')\hat{\bf L}_{km}$ after Fourier transformation
$\int d\omega e^{i\omega t}F(t)$ yields the retarded(advanced)
self-energy matrix $\hat{\bf \Sigma}^{r/a}_L(\omega)$  Eq. (14).

The procedure to get the retarded(advanced) Green's functions for
the superconductor lead is similar. After the following
Bogoliubov-Valatin transformation
\begin{eqnarray}
s_{p\uparrow}&=&\mu_p b_{p\uparrow}+\nu_p b_{-p\downarrow}^{\dagger},\\
s_{-p\downarrow}&=&\mu_p b_{-p\downarrow}-\nu_p
b_{p\downarrow}^{\dagger},
\end{eqnarray}
with
\begin{eqnarray}
\mu_p^2&=&\frac{1}{2}(1+\frac{\epsilon_{p\sigma}}
{\epsilon^2_{p\sigma}+\Delta^2}),\\
\nu_p^2&=&\frac{1}{2}(1-\frac{\epsilon_{p\sigma}}
{\epsilon^2_{p\sigma}+\Delta^2}),
\end{eqnarray}
the superconductor hamiltonian $H_S$ is diagonalized in the
following form
\begin{eqnarray}
 H'_S&=&\sum_p
\sqrt{\epsilon^2_{p\sigma}+\Delta^2}
(b^{\dagger}_{p\uparrow}b_{p\uparrow}+
b^{\dagger}_{-p\downarrow}b_{-p\downarrow})+2\sum_{p\sigma}
 \epsilon_{p\sigma} \nu_p^2-\nonumber \\ && \hspace{1cm}
 2\Delta \sum_p \mu_p \nu_p.
\end{eqnarray}
Using the notation
\begin{eqnarray}
g^{r/a}_{b\sigma;e}(t-t')&=&\mp i\theta(\pm t \mp t')\sum_{p}
 <{b_{p\sigma}(t),
b_{p\sigma}^{\dagger}(t')}>\nonumber \\
&=& \mp i\theta(\pm t \mp t') \sum_{p}e^{-i\sqrt{\epsilon
_{p\sigma}^2+\Delta^2}(t-t')
/\hbar}, \\
g^{r/a}_{b\sigma;h}(t-t')&=&\mp i\theta(\pm t \mp t')\sum_{p}
<{b_{p\sigma}
(t),b_{p\sigma}^{\dagger}(t')}>\nonumber \\
&=& \mp i\theta(\pm t \mp t')\sum_{k} e^{i\sqrt{\epsilon
_{p\sigma}^2+\Delta^2} (t-t') /\hbar},
\end{eqnarray}
one can express the retarded (advanced) Green's function matrix
for the superconductor lead as
\begin{equation}
\hat {\bf g}_{ss}^{r/a}(t-t')=\sum_{p} \hat{\bf
g}^{r/a}_{pp}(t-t') =(\hat {g}_{ss;mn}^{r/a}(t-t')),
\end{equation}
where
\begin{eqnarray}
\hat {g}_{ss;11}^{r/a}(t-t')&=&\mu^2_pg^{r/a}
_{b\uparrow;e}(t-t')+
\nu^2_pg^{r/a}_{b\downarrow;h}(t-t'),\\
\hat {g}_{ss;22}^{r/a}(t-t')&=&\mu^2_pg^{r/a}
_{b\downarrow;h}(t-t')+
\nu^2_pg^{r/a}_{b\uparrow;e}(t-t'),\\
\hat {g}_{ss;33}^{r/a}(t-t')&=&\mu^2_pg^{r/a}
_{b\downarrow;e}(t-t')+
\nu^2_pg^{r/a}_{b\uparrow;h}(t-t'),\\
\hat {g}_{ss;44}^{r/a}(t-t')&=&\mu^2_pg^{r/a}
_{b\uparrow;h}(t-t')+
\nu^2_pg^{r/a}_{b\downarrow;e}(t-t'),\\
\hat {g}_{ss;12}^{r/a}(t-t')&=&\hat {g}_{ss;21}^{r/a}(t-t')=\mu_p \nu_p
(g^{r/a}_{b\downarrow;h}(t-t')\nonumber \\ &&-g^{r/a}_{b\uparrow;e}(t-t')) ,\\
\hat {g}_{ss;34}^{r/a}(t-t')&=&\hat {g}_{ss;43}^{r/a}(t-t')=\mu_p \nu_p
(g^{r/a}_{b\downarrow;e}(t-t')\nonumber \\ &&-g^{r/a}_{b\uparrow;h}(t-t')) ,\\
\hat {g}_{ss;13}^{r/a}(t-t')&=&\hat {g}_{ss;14}^{r/a}(t-t') =\hat
{g}_{ss;23}^{r/a}(t-t')\nonumber \\ &=&
\hat {g}_{ss;24}^{r/a}(t-t')=0,\\
\hat {g}_{ss;31}^{r/a}(t-t')&=&\hat {g}_{ss;32}^{r/a}(t-t')= \hat
{g}_{ss;41}^{r/a}(t-t')\nonumber \\ &=&\hat
{g}_{ss;42}^{r/a}(t-t')=0.
\end{eqnarray}
Since
\begin{eqnarray}
\int d\epsilon_p(&\mu_p^2& e^{-i\sqrt{\epsilon_p^2+\Delta^2}\tau/\hbar}+
\nu_p^2 e^{i\sqrt{\epsilon_p^2+\Delta^2}\tau/\hbar}) \nonumber \\
&=&\frac12 \int_{-\infty}^{\infty}
d\epsilon_p(e^{-i\sqrt{\epsilon_p^2+\Delta^2}\tau/\hbar}+\nonumber
\\ & &\hspace{1cm}
 e^{i\sqrt{\epsilon_p^2+\Delta^2}\tau/\hbar}) \nonumber\\
&=& \int_0^{\infty} d \epsilon \frac{\epsilon \theta(\epsilon-\Delta)}
{\sqrt{\epsilon^2-\Delta^2}}(e^{-i\epsilon\tau/\hbar}+
 e^{i\epsilon\tau/\hbar})  \nonumber \\
&=& \int_{-\infty}^{\infty}d\epsilon
\frac{|\epsilon| \theta(|\epsilon|-\Delta)}
{\sqrt{\epsilon^2-\Delta^2}}e^{-i\epsilon\tau/\hbar},
\end{eqnarray}
and
\begin{eqnarray}
\int d\epsilon_p\mu_p& \nu_p&
(e^{-i\sqrt{\epsilon_p^2+\Delta^2}\tau/\hbar}-
e^{i\sqrt{\epsilon_p^2+\Delta^2}\tau/\hbar})\nonumber \\
&=&\frac12 \int_{-\infty}^{\infty} d\epsilon_p
\frac{\Delta}{\sqrt{\epsilon_p^2
+\Delta^2}}(e^{-i\sqrt{\epsilon_p^2+\Delta^2}\tau/\hbar}-\nonumber
\\ & &\hspace{1cm}
 e^{i\sqrt{\epsilon_p^2+\Delta^2}\tau/\hbar}) \nonumber \\
&=& \int_0^{\infty} d \epsilon \frac{\Delta \theta(\epsilon-\Delta)}
{\sqrt{\epsilon^2-\Delta^2}}(e^{-i\epsilon\tau/\hbar}-
 e^{i\epsilon\tau/\hbar})  \nonumber \\
&=& -\int_{-\infty}^{\infty}d\epsilon
\frac{|\epsilon| \theta(|\epsilon|-\Delta)}
{\sqrt{\epsilon^2-\Delta^2}}\frac{\Delta}{\epsilon}
e^{-i\epsilon\tau/\hbar},
\end{eqnarray}
then one obtains after changing the sum $\sum_p$ into an integral
$\int d\epsilon_p \rho_S^N$
\begin{widetext}
\begin{eqnarray}
\hat {g}_{ss;11}^{r/a}(t-t')&=&\hat {g}_{ss;22}^{r/a}(t-t')
=\hat {g}_{ss;33}^{r/a}(t-t')=\hat {g}_{ss;44}^{r/a}(t-t')\\
&=& \mp i\theta(\pm t \mp t') \int d\epsilon
\frac{|\epsilon| \theta(|\epsilon|-\Delta)}
{\sqrt{\epsilon^2-\Delta^2}}e^{-i\epsilon(t-t')/\hbar} \\
\hat {g}_{ss;12}^{r/a}(t-t')&=&\hat {g}_{ss;21}^{r/a}(t-t')
=-\hat {g}_{ss;34}^{r/a}(t-t')=-\hat {g}_{ss;43}^{r/a}(t-t')\\
&=& \mp i\theta(\pm t \mp t') \int d\epsilon
\frac{|\epsilon| \theta(|\epsilon|-\Delta)}
{\sqrt{\epsilon^2-\Delta^2}}\frac{\Delta}{\epsilon}
e^{-i\epsilon(t-t')/\hbar}, \\
\hat {g}_{ss;13}^{r/a}(t-t')&=&\hat {g}_{ss;14}^{r/a}(t-t')
=\hat {g}_{ss;23}^{r/a}(t-t')=
\hat {g}_{ss;24}^{r/a}(t-t')=0,\\
\hat {g}_{ss;31}^{r/a}(t-t')&=&\hat {g}_{ss;32}^{r/a}(t-t')= \hat
{g}_{ss;41}^{r/a}(t-t')=\hat {g}_{ss;42}^{r/a}(t-t')=0.
\end{eqnarray}
\end{widetext}
Hence the retarded(advanced) self-energy matrix $\hat{\bf
\Sigma}^{r/a}_R(\omega)$  Eq. (15) is similarly obtained from the
direct matrix product $\hat{\bf R}_{pn}^\dagger \hat{\bf
g}_{ss}^{r/a}(t-t')\hat{\bf R}_{pm}$ after Fourier transformation
$\int d\omega e^{i\omega t}F(t)$.


\newpage

\begin {thebibliography}{30}
\bibitem {Kouwenhoven} For a review, see  {\it
Mesoscopic Electronic Transport}, Edited by L. L. Sohn, L. P.
Kouwenhoven, and G. Sch{\"o}n, (Kluwer, Series E 345, 1997).
\bibitem {Kittel} C. Kittel, {\it Introduction to Solid State Physics}
           ( Willy and Sons, New York, 1976).
\bibitem {Kubo} R. Kubo, M. Toda, and N. Hasnitsume, {\it Nonequilibrium
Statistical Mechanics} (Springer-Verlag, Berlin, 1997).
\bibitem {Datta} S. Datta, {\it Electronic Transport in Mesoscopic Systems}
(Cambridge University Press, 1995), P246-273.
\bibitem {Buttiker} R. Landauer, Philos.
Mag. {\bf 21}, 863 (1970); M. B\"uttiker, Y. Imry, R. Landauer,
and S. Pinhas, Phys. Rev. B {\bf 31}, 6207 (1985).
\bibitem {Poirier} W. Poirier, D. Mailly, and M. Sanquer, Phys. Rev. Lett.
 {\bf 79}, 2105 (1997).
\bibitem {Post} N. van der Post, E. T. Peters, I. K. Yanson,
and J. M. van Ruitenbeek, Phys. Rev. Lett. {\bf 73}, 2611 (1994).
\bibitem {Morpurgo}A. F. Morpurgo, B. J. van Wees, T. M. Klapwijk,
and G. Borghs, Phys. Rev. Lett. {\bf 79}, 4010 (1997).
\bibitem {Tuominen} M. T. Tuominen, J. M. Hergenrother, T. S. Tighe,
 and M. Tinkham, Phys. Rev. Lett. {\bf 69}, 1997 (1992).
\bibitem {Eiles} T. M. Eiles, John M. Martinis, and Michel H.
Devoret, Phys. Rev. Lett. {\bf 70}, 1862 (1993). Shashi K.
\bibitem {Upadhyay} S. K. Upadhyay, A. Palanisami, R. N. Louie, and R. A.
Buhrman, Phys. Rev. Lett. {\bf 81}, 3247 (1999).
\bibitem {Lawrence} M. D. Lawrence and N. Giordano, J. Phys. Condens. Matter {\bf 39},
L563(1996).
\bibitem  {Gueron} S. Gueron, Mandar M. Deshmukh, E. B. Myers,
 and D. C. Ralph, Phys. Rev. Lett. {\bf 83}, 4148 (1999).
\bibitem {Andreev} A. F. Andreev, Sov. Phys. JETP {\bf 19}, 1228
 (1964).
 \bibitem {Prinz} G. A. Prinz, Science {\bf 282}, 1660 (1998).
 \bibitem {Jong} M. J. M. de Jong and C. W. J. Beenakker, Phys. Rev. Lett. {\bf 74},
 1657 (1999).
 \bibitem {Haug} H. Haug and A. -P. Jauho, {\it Quantum Kinetics
 in Transport and Optics of Semiconductors} (Springer-Verlag,
 Berlin, 1998).
 \bibitem {Meir} Y. Meir and N. S. Wingreen, Phys. Rev. Lett. {\bf 68}, 2512
 (1992).
 \bibitem {Sun} Qing-feng Sun, Jian Wang, and Tsung-han Lin,
 Phys. Rev. B {\bf 59}, 3831 (1999).
 \bibitem {Yeyati}
 A. Levy Yeyati, J. C. Cuevas, A. López¡¡Dávalos, and A.
 Martn¡¡Rodero, Phys. Rev. B {\bf 55}, R6137 (1997).
 \bibitem {Slonczewski} J. C. Slonczewski, Phys. Rev. B {\bf 39}, 6995
 (1989).
 \bibitem {Gennes} P. G. de Gennes, {\it Superconductivity of
 Metals and Alloys} (Benjamin, New York, 1966).
\bibitem {Jauho} A. P. Jauho and N. S. Wingreen, Phys. Rev. B {\bf 50},
 5528 (1994).
\bibitem {Bogoliubov} N. N. Bogoliubov, Nuovo Cimento {\bf 7},
 794 (1958); J. G. Valatin, Nuovo Cimento {\bf 7}, 843 (1958).

\end {thebibliography}
\end{document}